\documentstyle[aps,prl,preprint,floats,epsfig]{revtex}

\begin{document}

\preprint{\tighten\vbox{\hbox{\hfil CLEO CONF 01-03}
                        \hbox{\hfil EPS 853}
                        \hbox{\hfil Lepton Photon 543}
}}

\title{Measurement of the $\Xi^{+}_{c}$ Lifetime}

\author{CLEO Collaboration}
\date{July 13, 2001}

\maketitle
\tighten

\begin{abstract} 
The $\Xi^{+}_{c}$ lifetime is measured using 9.0~fb$^{-1}$ of $e^+e^-$
annihilation data collected on and just below the $\Upsilon(4S)$
resonance with the CLEO II.V detector at CESR.  The $\Xi^{+}_{c}$
lifetime is measured using an unbinned maximum likelihood fit.  The
preliminary result for the $\Xi^{+}_{c}$ lifetime is 
$(503 \pm 47 ({\rm stat.}) \pm 18 ({\rm syst.}))$ fs.
\end{abstract}
\vfill
\begin{flushleft}
.\dotfill .
\end{flushleft}
\begin{center}
Submitted to the \\
International Europhysics Conference on High Energy Physics \\
EPS HEP-2001 July 2001, Budapest Hungary \\ 
and the \\
XXth International Symposium on Lepton and Photon Interactions at High
Energies\\  
Lepton-Photon July 2001, Rome Italy
\end{center}
\newpage

{
\renewcommand{\thefootnote}{\fnsymbol{footnote}}

\begin{center}
M.~Artuso,$^{1}$ C.~Boulahouache,$^{1}$ K.~Bukin,$^{1}$
E.~Dambasuren,$^{1}$ G.~Majumder,$^{1}$ R.~Mountain,$^{1}$
T.~Skwarnicki,$^{1}$ S.~Stone,$^{1}$ J.C.~Wang,$^{1}$
H.~Zhao,$^{1}$
% ...
\end{center}
 
\small
\begin{center}
$^{1}${Syracuse University, Syracuse, New York 13244}\\
% ...
\end{center}

\setcounter{footnote}{0}
}
\newpage

\section{Introduction}

  Charm baryon lifetime measurements provide insight into the 
dynamics of non-perturbative heavy quark decays.  The theoretical
situation is rich with possibilities. Unlike the case of charm mesons
 the exchange mechanism  is not  helicity suppressed and therefore can be 
comparable in magnitude to the spectator diagram. In addition, color suppression is only active
for particular decay channels. Thus spectator decays alone can not
account for the hadronic width in charm baryon decay. 
The hadronic width  
is modified by at least three effects: (a) destructive interference between 
external and internal
spectator diagrams, (b) constructive interference between internal spectator diagrams,
and (c) $W$-exchange diagrams. Effects (a) and (b) are expected to 
be operative in the decay
of the $\Xi^{+}_{c}$, while (a) and (c) play a role in $\Lambda^{+}_{c}$ decay. 
While several models~\cite{BBD} can account for the apparent lifetime hierarchy,
$\tau_{\Xi^{+}_{c}}>\tau_{\Lambda^{+}_{c}}>\tau_{\Xi^{0}_{c}}>\tau_{\Omega_{c}}$,
a detailed understanding of the  various contributions to the hadronic width 
requires input from experiment. 

The lifetimes of the charm baryons are not measured as precisely as the charm
mesons ($D^0$, $D^+$, $D_s$) which 
are measured, by individual experiments~\cite{pdg}, to a  
precision of $\sim$1 - 3\%.  
The $\Lambda^{+}_{c}$'s lifetime
is the most precisely measured of the charm baryons. Recently, 
 CLEO and SELEX measured this lifetime to  a precision of 5\%~\cite{lbc_cleo,lbc_selex}.  
Other charm baryon lifetimes ($\Xi^{+}_{c}$, $\Xi^{0}_{c}$, and $\Omega_{c}$) 
are measured to $\ge$ 20\% uncertainty~\cite{pdg}.  
This paper presents 
CLEO's preliminary measurement of the $\Xi^{+}_{c}$ lifetime, the first measurement
of this lifetime from an $e^+e^-$ colliding beam experiment.  

\section{Data Sample and Event Selection}
This analysis uses an integrated luminosity
of 9.0 fb$^{-1}$ of $e^+e^-$ annihilation data taken with the CLEO II.V detector
at the Cornell Electron Storage Ring (CESR).  The data were taken at energies at and
below the $\Upsilon$(4S) resonance ($\sqrt{s}$ =10.58 GeV) and include
$\sim$ 11$\times 10^6~e^+e^-\to c \bar{c}$ events.
This analysis relies heavily on the charged particle tracking
capabilities of the CLEO II.V detector~\cite{cleod}. 
The precise location of both primary
and secondary vertices is greatly aided by a small-radius, low-mass beam pipe
surrounded by a three-layer double-sided silicon strip tracker~\cite{svx}.  The trajectories
of charged particles are reconstructed using two drift chamber systems
 in addition to
the silicon strip tracker. For this data set the main drift chamber uses a 60:40 mixture of helium propane
gas in place of its standard 50:50 argon-ethane mix. This change in gas improves both 
the hit efficiency and specific ionization resolution
while at the same time decreases the effects of multiple scattering. A Kalman filter 
 track reconstruction algorithm~\cite{KALMAN} is used to reconstruct 
the three dimensional trajectories
of charged particles. The response of the detector to both signal and
background events is modeled in detail using the GEANT~\cite{GEANT} Monte Carlo package.
 
  The $\Xi^{+}_{c}$ is reconstructed from the decay $\Xi^{-} \pi^{+} \pi^{+}$.  Each
$\Xi^{-}$ is reconstructed using the $\Lambda \pi^{-}$ mode while $\Lambda$'s are reconstructed from
$p \pi^{-}$ (the charge conjugate mode is implied throughout this paper).  For this analysis we
assume that the $\Xi^{+}_{c}$ is produced at the primary event vertex and is not a decay product of
another long lived weakly decaying particle.    
Each $\Xi^{+}_{c}$ candidate's proper time, $t$, and 
proper time uncertainty, $\sigma_{t}$,
is determined from 
\begin{equation}
  t = \frac{m_{\Xi^{+}_{c}}}{c p_{y_{\Xi^{+}_{c}}}} (y_{\rm decay} - y_{\rm production})
\end{equation}
and
\begin{equation}
  \sigma_{t} = \frac{m_{\Xi^{+}_{c}}}{c \mid p_{y_{\Xi^{+}_{c}}} \mid} 
  \sqrt{\sigma^{2}_{y_{\rm decay}} + \sigma^{2}_{y_{\rm beam~position}} + 
        \sigma^{2}_{y_{\rm beam~size}}}
  \hspace{0.05 in}.
\end{equation}
In the above equations
the $\Xi^{+}_{c}$ mass, $m_{\Xi^{+}_{c}}$, is set to the PDG~\cite{pdg} 
value of 2466.3$\pm1.4$ MeV/${\rm c^{2}}$.
A best-fit decay vertex of the $\Xi^{-}$ pseudo-track and
the two $\pi^{+}$'s determines $y_{\rm decay}$
and its uncertainty, $\sigma_{\rm y_{decay}}$. 
The production point of a $\Xi^{+}_{c}$ cannot be well measured on an event-by-event
basis. Instead we use a combination of the known CESR beam profile and a measurement of the beam
centroid to provide the estimate of the $\Xi^{+}_{c}$ production point.
In the CLEO environment the dimensions ($\sigma$) of the beam profile are 1 cm along the
beam line, $z$, 350 $\mu$m along the horizontal direction perpendicular to the beam line, $x$,
and 7 $\mu$m ($\sigma_{y_{\rm beam~size}}$) along the vertical direction, $y$. 
Since the typical decay length of 
a $\Xi^{+}_{c}$ ($\sim 150$ $\mu$m) is significantly less 
than the beam extent in $z$ and $x$ essentially all
useful decay length information comes from the $y$-coordinate.   
A run-averaged collision point of 
the $e^{+}$ and $e^{-}$ beams, $y_{\rm beam~position}$,
is used to estimate $y_{\rm production}$.
The uncertainty in $y_{\rm beam~position }$, $\sigma_{y_{\rm beam~position}}$,
is also calculated run-by-run.
The $\Xi^{+}_{c}$'s 
component of momentum in the $y$ direction, $ p_{y_{\Xi^{+}_{c}}}$, 
is calculated 
from the momentum of its decay products.  
     
\begin{figure}[!htb]
          \begin{center}

            \epsfig{file=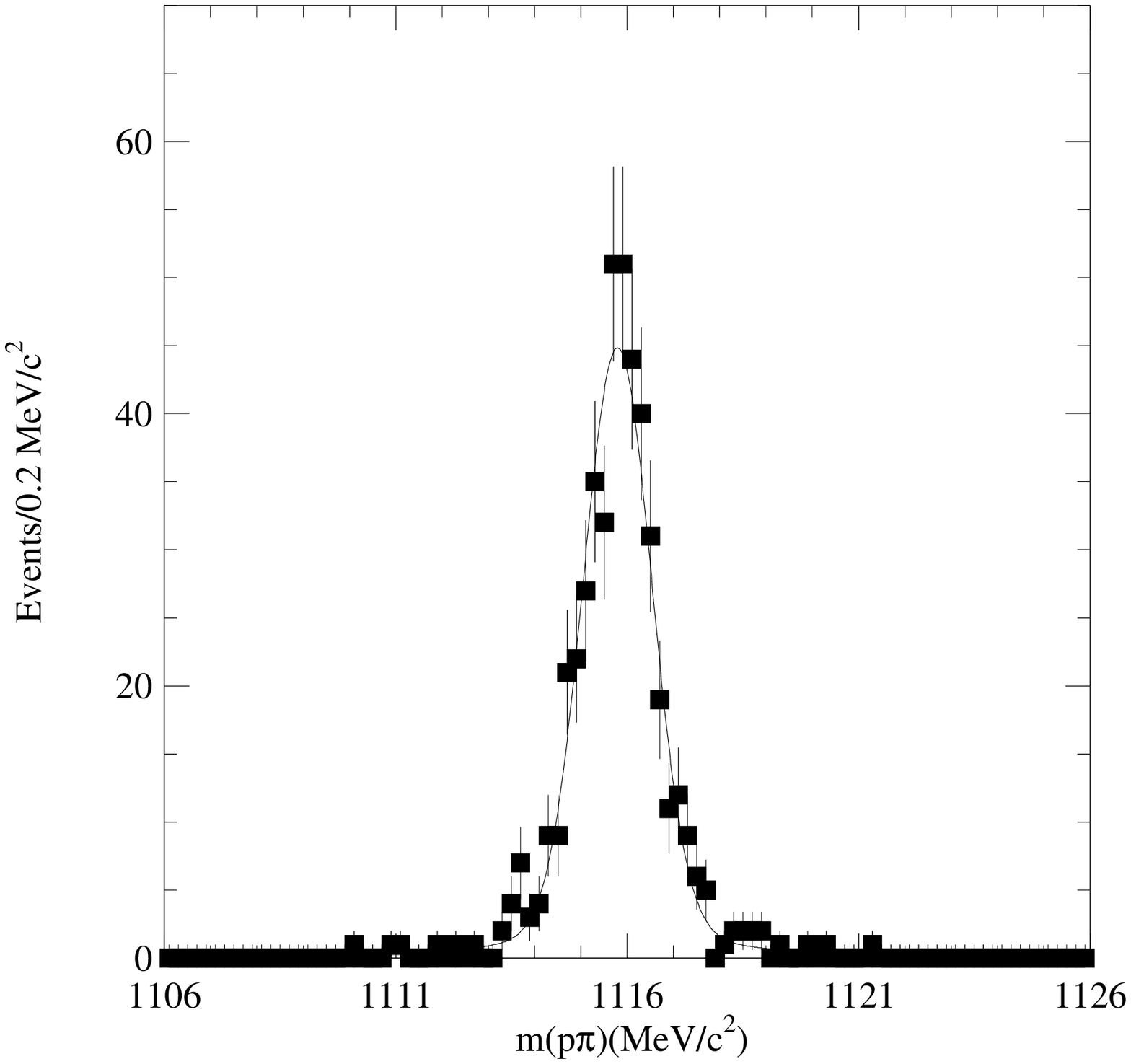,width=2.0in,
                    bbllx=12bp,bblly=15bp,bburx=526bp,bbury=495bp,clip= } 
            \epsfig{file=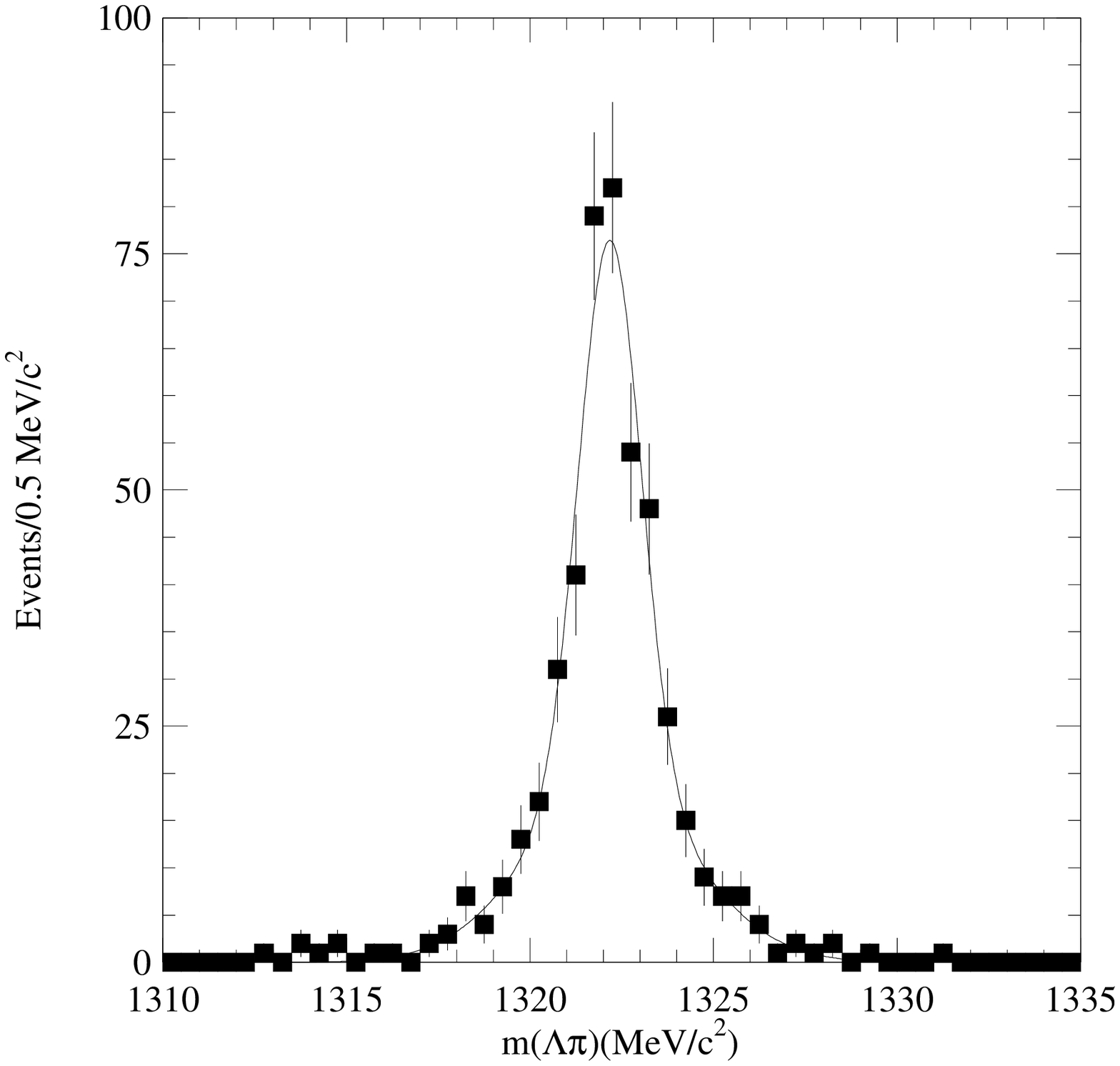,width=2.0in,
                    bbllx=12bp,bblly=15bp,bburx=525bp,bbury=500bp,clip= } 
            \epsfig{file=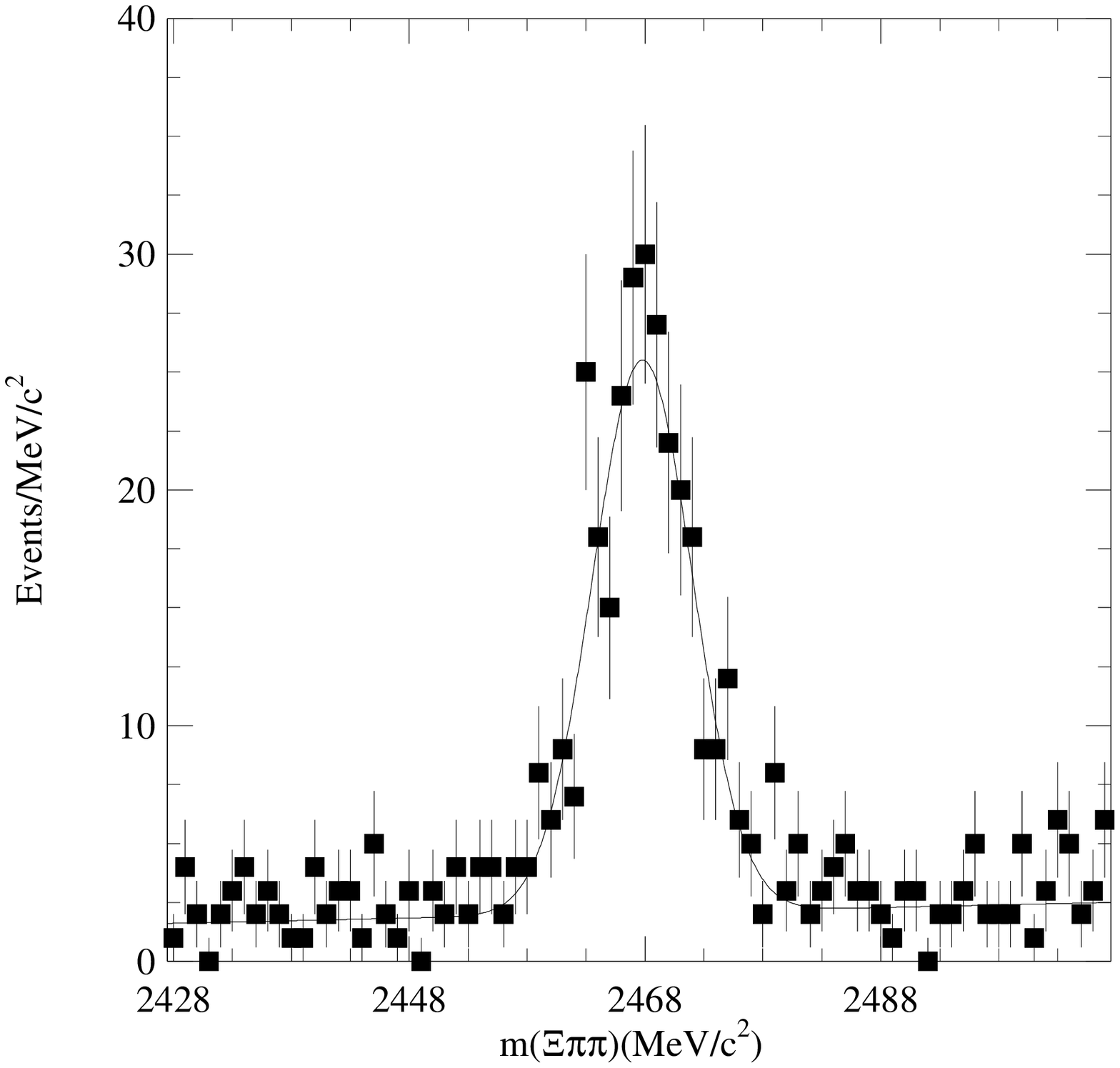,width=3.0in,
                    bbllx=12bp,bblly=15bp,bburx=525bp,bbury=500bp,clip= }  
            \caption{Invariant mass distributions of the  $\Lambda \rightarrow p \pi^{-}$, 
		     $\Xi^{-} \rightarrow \Lambda \pi^{-}$, and 
		     $\Xi^{+}_{c} \rightarrow \Xi^{-} \pi^{+} \pi^{+}$ 
		     candidates used to determine the $\Xi^{+}_{c}$ lifetime.
              }
            \label{fig:ximass}
          \end{center}
        \end{figure}

While much of the $\Xi^{+}_{c}$ selection criteria are similar to that of previous CLEO 
charm baryon analyses~\cite{cleo_Xi0p,cleo_Xih,cleo_omegac} 
some additional requirements suited for a lifetime measurement are imposed.
To select high momentum  candidates and reduce $B$-meson related backgrounds
we require  the $\Xi^{+}_{c}$'s 
momentum fraction,  $p_{\Xi^{+}_{c}}/p_{\Xi^+_c~{\rm max}}$, 
 to be larger than 0.5.  For each candidate
we impose $ \sigma_{t}<$ 1.5 ps.  A minimum $\Xi^{-} \pi^{+} \pi^{+}$ vertex probability of 0.001 
(based on the vertex $\chi^{2}$) 
is required to obtain a sample of well-defined decay lengths.  To ensure that only
one $\Xi^{+}_{c}$ candidate per event is used, the candidate with the smallest
vertex $\chi^{2}$ is chosen in the events where multiple candidates pass all
other selection criteria ($\sim$ 7\% of events).  Figure~\ref{fig:ximass} 
shows the $\Lambda$, $\Xi^{-}$,
and $\Xi^{+}_{c}$ reconstructed mass distributions for candidates
used in the lifetime analysis.
A fit of the $\Xi^{+}_{c}$ mass distribution using one Gaussian for the signal and a linear 
function for the background yields $250 \pm 18$ reconstructed $\Xi^{+}_{c}$'s
and a Gaussian $\sigma$ of 4.3 MeV/c$^2$.
The fraction of background within $\pm 2~\sigma$  of the 
fitted $\Xi^{+}_{c}$ mass is 12.8\%.
The average flight path in the $y$ direction
of the $\Xi^{+}_{c}$'s used in this analysis is 100~$\mu$m. 
The efficiency of the selection cuts for detecting signal Monte Carlo events, including
(not including)  acceptance, is  7.6\% (17\%). 
Events within $\pm$40 MeV/${\rm c^{2}}$ of the mean
reconstructed $\Xi^{+}_{c}$ mass (2468 MeV/${\rm c^{2}}$), as shown in 
Figure~\ref{fig:ximass}, are used in the determination of the $\Xi^{+}_{c}$ lifetime.
This wide mass region is used to estimate the non-$\Xi^{+}_{c}$ contribution to the lifetime.

\section{Lifetime Determination Method} 

  The $\Xi^{+}_{c}$ lifetime is obtained from an unbinned maximum likelihood
fit to the proper time distribution. 
 The likelihood function is 
\begin{eqnarray*}
  \lefteqn{L(\tau_{\rm sig} , S, \sigma_{\rm 
    mis}, f_{\rm mis}, \tau_{\rm BG} , f_{\tau_{BG}} , f_{\rm flat}) = } \\ 
       &    &    \prod_i \int_0^\infty
  \left[
    \underbrace{p_{\rm sig,i}E(t^\prime|\tau_{\rm sig})}_{\rm signal\ fraction} +\right.
  \left.
    \underbrace{(1-p_{\rm sig,i})\left(f_{\tau_{\rm BG}} E(t^\prime|\tau_{\rm BG}) +
                        (1-f_{\tau_{\rm BG}})\delta(t^\prime)\right)}_{\rm
    background\ fraction}
  \right] \nonumber \\
    &    &    \times \left[
    \underbrace{(1-f_{\rm mis}-f_{\rm flat})G(t_{\rm i}-t^\prime|S \sigma_{\rm t,i})}_{\rm
    proper\ time\
    resolution} +
       \underbrace{f_{\rm mis}G(t_{\rm i}-t^\prime|\sigma_{\rm
    mis})}_{\rm mis-measured\ frac.} +
       \underbrace{f_{\rm flat}G(t_{\rm i}-t^\prime|\sigma_{\rm
    flat})}_{\rm flat\ frac.}
  \right]
  dt^\prime \nonumber
\end{eqnarray*}

\noindent
with the product over all $\Xi^{+}_{c}$ candidates, 
$G(t|\sigma)\equiv \exp(-t^2/2\sigma^2)/\sqrt{2\pi}\sigma$
and $E(t|\tau) \equiv \exp(-t/\tau)/\tau$.  

There are three inputs to the fit for each  
 $\Xi^{+}_{c}$ candidate: the measured proper time, $t_{\rm i}$, the estimated uncertainty
in the proper time, $\sigma_{\rm t,i}$, and a mass dependent signal probability, $p_{\rm sig,i}$.
The signal probability distribution is obtained from a fit to the  $\Xi^{+}_{c}$ mass distribution.

The proper time distribution is
parameterized as consisting of signal events with lifetime, $\tau_{\rm sig}$, a fraction,
$f_{\tau_{\rm BG}}$, of
background events with non-zero lifetime, $\tau_{\rm BG}$, from charm backgrounds, 
and the remaining background events with zero lifetime.  
The likelihood function allows for a global scale factor, $S$, for the proper time
uncertainties.  The likelihood function also accounts for events in which the
proper time uncertainty is underestimated by fitting for a $\sigma_{\rm mis}$ 
and fraction, $f_{\rm mis}$ (caused by non-Gaussian multiple scattering, for instance) and 
also a fraction, $f_{\rm flat}$ of events with a fixed $\sigma_{\rm flat}$ = 8 ps to
account for proper time outliers. 

\begin{figure}[!htb]
          \begin{center}

            \epsfig{file=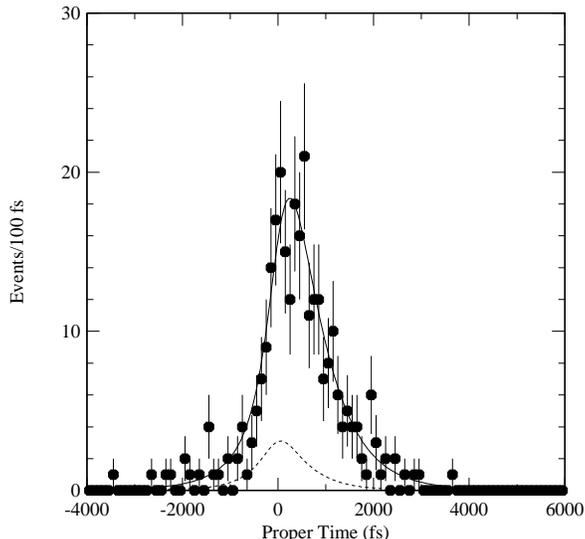,width=3.0in,
                    bbllx=16bp,bblly=11bp,bburx=526bp,bbury=487bp,clip= } 
            \caption{Proper time distribution of events within $\pm$2 $\sigma$ of 
		     the $\Xi^{+}_{c}$ mass peak.
		     The scaled proper time fit (solid line)
	             and scaled background component of the fit (dotted line) are superimposed on the data.}
            \label{fig:xicpropertimedist}
          \end{center}
        \end{figure}

The unbinned maximum likelihood fit yields a
signal lifetime, $\tau_{\rm sig}= 497 \pm 47$ fs.  The proper time distribution and the unbinned
maximum likelihood fit are shown in Figure~\ref{fig:xicpropertimedist}.

In order to check the consistency of this lifetime result the analysis procedure
is repeated as a function of  $\Xi^{+}_{c}$ charge, azimuthal angle,
polar angle, momentum, silicon detector hit criteria, and data taking
period. In all cases the lifetimes are statistically consistent.

        \section{Systematic Errors}
The contributions to the systematic error are given in Table~\ref{tab:mlsystematic}
and discussed below.

        \begin{table}[!htb]
          \begin{center} 		
	    \begin{tabular}{|l|c|} \hline
   		Contribution  &Uncertainty (fs) \\ \hline
                $\Xi^{+}_{c}$ mass & $\pm 0.3$ \\
                $\Xi^{+}_{c}$ momentum & $^{+0.6}_{-0.0}$  \\
                global detector size \& decay length bias& $\pm 2.5$ \\
                signal probability & $^{+2.7}_{-3.6}$ \\
                proper time outliers & $\pm 3.3$  \\
		$y$ beam position   & $\pm 7$ \\
                proper time - mass correlation & $\pm 7.2$\\
                Monte Carlo statistics & $\pm 9.2$ \\
                fit mass region & $\pm 10$ \\ \hline
		Total                  & $\pm 18$ \\ \hline
            \end{tabular}
            \caption{Contributions to the systematic error of the $\Xi^{+}_{c}$ lifetime.}
            \label{tab:mlsystematic}
          \end{center}
        \end{table}

The uncertainty of the $\Xi^{+}_{c}$ mass could be a source of
	bias in the lifetime measurement as this mass is used to 
	determine the proper time.  The PDG~\cite{pdg} uncertainty of the
	$\Xi^{+}_{c}$ mass of $\pm 1.4~\rm{MeV/c^{2}}$ yields a
	systematic error contribution of $\pm 0.3$ fs.
 The systematic bias of the $\Xi^{+}_{c}$ momentum from an
	incorrect magnetic field could yield a systematic shift 
	in the reconstructed masses.  Such a shift would then cause
	a bias in the lifetime measurement.  
We estimate this systematic
	 error contribution to be $^{+0.6}_{-0.0}$ fs.

The global detector size and beam
	pipe geometry is studied 
	to understand their contribution to the
systematic error in the lifetime. The results of the study
yield a lifetime uncertainty 
	of 0.1\% resulting in a systematic error contribution of
	$\pm$0.5 fs.  The potential bias in the decay length measurement is determined
	by measuring the average decay length of a ``zero-lifetime'' 
        sample of $\gamma \gamma \rightarrow 4 \pi$ events.  We measure 
	an average decay length of $0.0\pm$0.9 $\mu$m and 
        use the uncertainty in this measurement to calculate
        the contribution to the total systematic error. 
        The average of the quotient of 0.9 $\rm{\mu m}$
	and the $\beta \gamma c$ for each $\Xi^{+}_{c}$ candidate is
	2.5 fs. We take this to be the estimate of the $\Xi^{+}_{c}$ 
	proper time bias due to a decay length bias.  
	Adding these two systematic errors in quadrature yields 2.5 fs.

The signal probability, $p_{\rm sig,i}$, contribution to the systematic error
is obtained from differences
	in the fitted lifetime values when the signal probability
	 is varied by $\pm 1~\sigma_{p_{\rm sig,i}}$.  
	This study yields a systematic error of $^{+2.7}_{-3.6}$ fs.

	  The proper time outlier contribution is obtained from the
	maximal difference of lifetimes from the following three methods
	of accounting for outliers:
	a) a $\sigma_{\rm flat}$ = 8 ps contribution in the likelihood
	  function (this is the nominal method of accounting for 
	  proper time outliers),
	b) a $\sigma_{\rm flat}$ = 16 ps contribution in the likelihood
	  function, and c) a  proper time cut (absolute value) of less than 4 ps and no
	  $\sigma_{\rm flat}$ contribution to the likelihood function.
	The maximal difference between these three methods is 3.3 fs which is taken as the 
	proper time outlier systematic error.

The $y$ beam position systematic 
	error estimates the variation in the lifetime
	when the $y$ beam position is shifted from its true position. 
        Shifting the beam spot location
	subsequently shifts decay length
	and proper time measurements.  For an infinite data sample
	in a perfectly isotropic detector, a shift in the
	$y$ beam position would not affect a lifetime measurement 
	 as it would average out to zero.  
        A possible lifetime bias can
	be estimated by measuring the lifetime after shifting the
	$y$ beam spot location.  
	The beam spot location is shifted by various amounts, and in the 
	vicinity of zero shift, the slope of the change in lifetime vs. the
	change in beam position location is 3.5 fs/$\rm{\mu m}$.  
	Multiplying this slope by $\pm$$2~\rm{\mu m}$, the uncertainty in 
	the $y$ beam position, yields a systematic error of $\pm$7 fs.

\begin{figure}[!htb]
          \begin{center}
            \epsfig{file=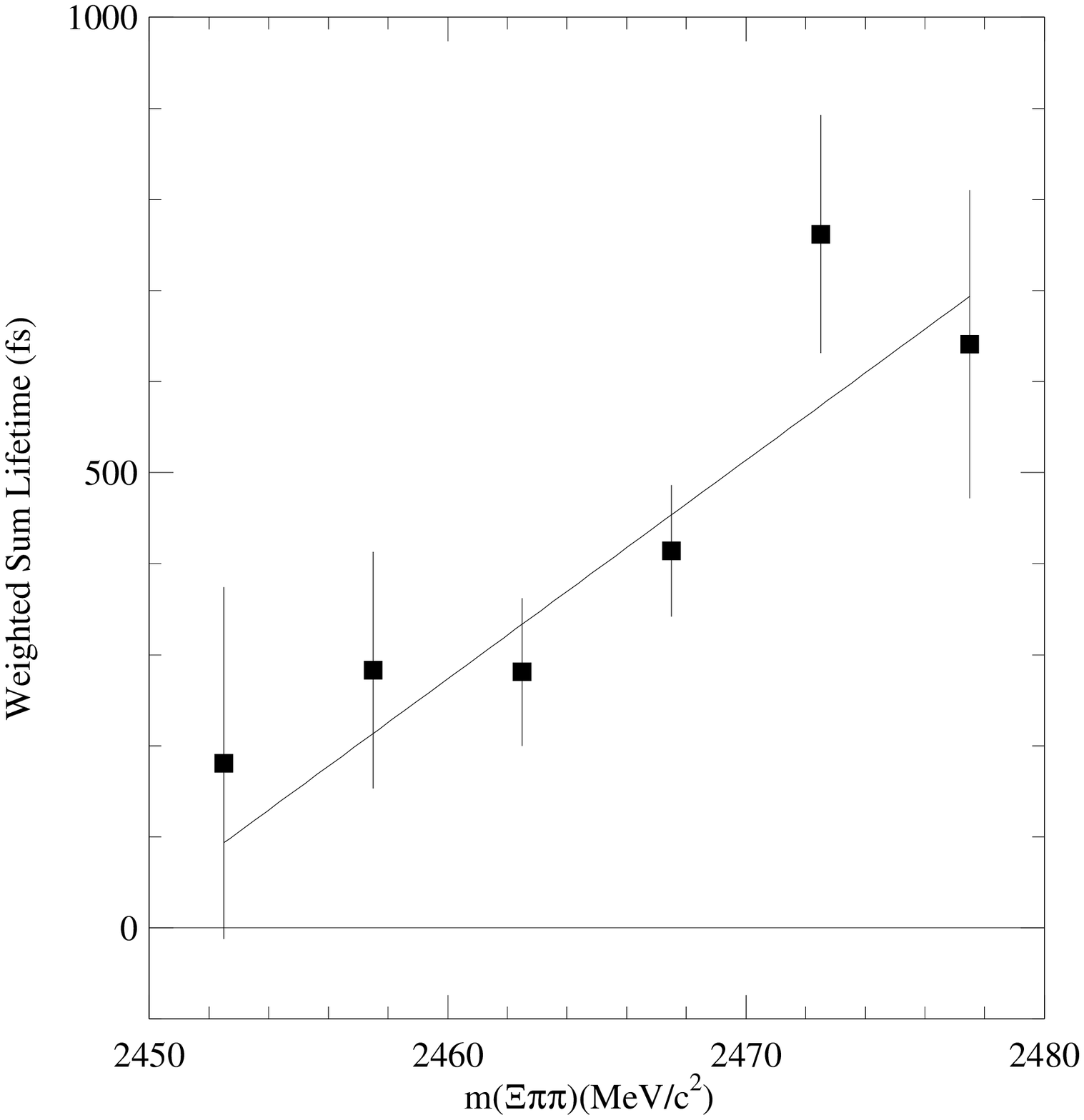,width=2.5in,
                    bbllx=16bp,bblly=15bp,bburx=526bp,bbury=538bp,clip= } 
            \epsfig{file=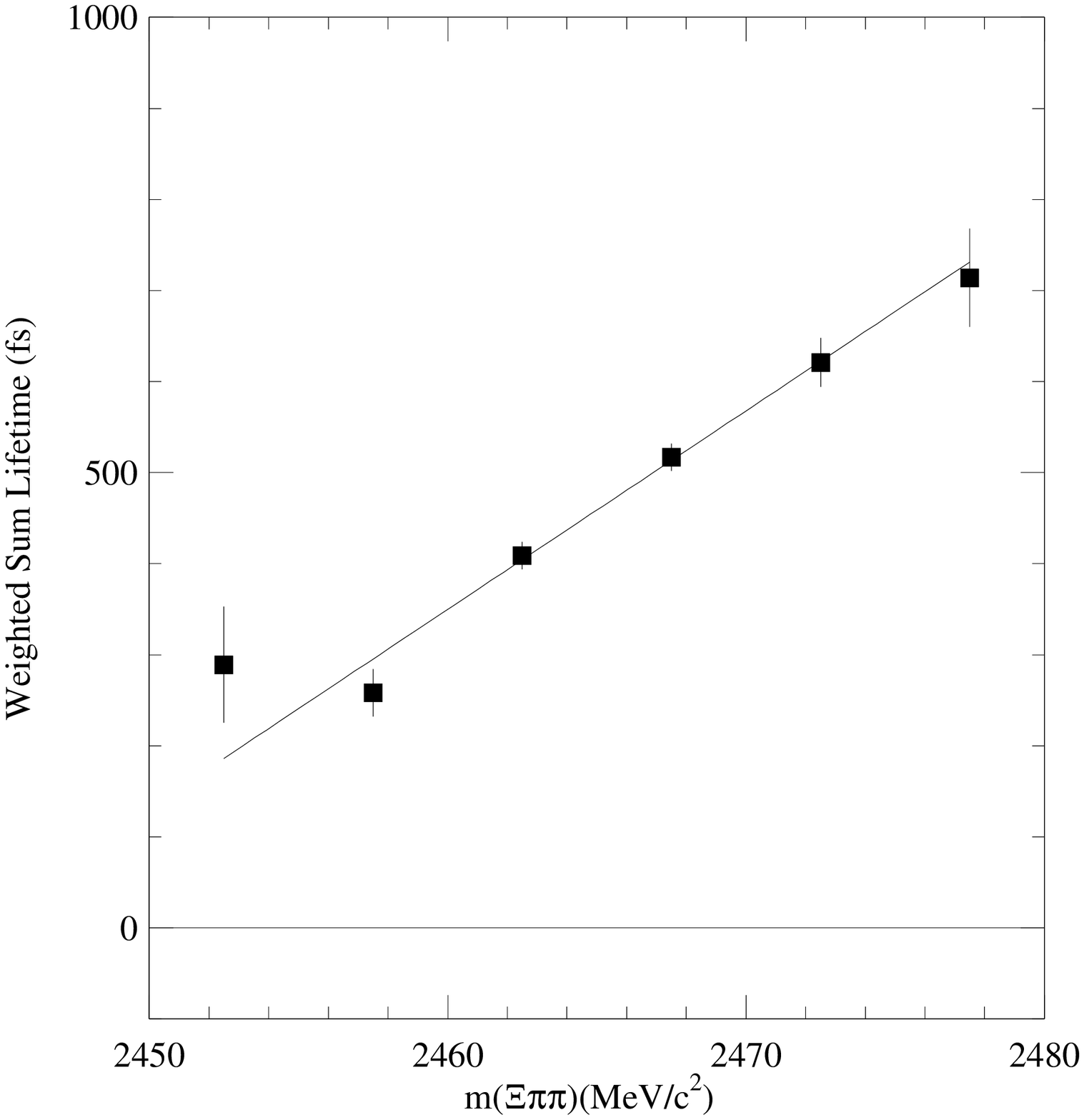,width=2.5in,
                    bbllx=16bp,bblly=15bp,bburx=526bp,bbury=538bp,clip= } 
            \caption{Average lifetime as a function of
	             invariant mass subregions near the signal region for 
		     CLEO II.V data (left) and tagged Monte Carlo events
		     (right).  For
		     data and Monte Carlo, the lifetime vs. invariant
		     mass slopes agree with values of 
		     24 and 22 $\rm{fs/(MeV/c^{2})}$, respectively.}
            \label{fig:mcdatatauvsivm}
          \end{center}
        \end{figure}

There is a correlation between the measured proper time and reconstructed
mass of a charm meson  or baryon. This correlation is due to 
the mis-measurement of the opening
angle(s) between the daughter tracks in a short-lived decay. 
An overestimate of the opening angle(s) tends to bias 
the measured proper time and mass to larger values of these quantities.
In Fig.~\ref{fig:mcdatatauvsivm} 
the average lifetime vs. invariant mass correlation is shown for case 
of the $\Xi^{+}_{c}$ using both data and Monte Carlo events.  
 There is good agreement in
the rate of change in lifetime vs. invariant mass 
for the two samples.
The proper time vs. mass  contribution to the systematic error
is obtained by multiplying the slope of the lifetime vs. measured $\Xi^{-}\pi^{+}\pi^{+}$ 
	reconstructed mass 
	by the $\sigma$ of the central mass value. 
This contribution to the systematic uncertainty is
$\pm$7.2 fs.

	In order to check for other sources that could bias
the lifetime measurement, e.g. event selection and
likelihood function parameterization,  a sample consisting of 
	background events extracted from CLEO II.V data and simulated $\Xi^{+}_{c}$ events 
	is studied. The relative amount of signal and background component 
in the sample is arranged to be the same as that of the full CLEO II.V
	data set.  This data set is run through the
full analysis  and the $\Xi^{+}_{c}$ lifetime extracted from the maximum likelihood fit
is compared with the input  lifetime (449.0 fs).   
This procedure yields a $\Xi^{+}_{c}$ lifetime
	of 443.2 $\pm$ 9.2 fs, 5.8 fs lower than the input Monte Carlo signal lifetime.	
	The statistical uncertainty in this measurement, 9.2 fs, is included as a
component of the total systematic error (``Monte Carlo statistics'') in 
Table~\ref{tab:mlsystematic}.  
	The 5.8 fs difference between
	the input and output signal Monte Carlo lifetime is applied 
as a correction to the $\Xi^{+}_{c}$ lifetime value from the CLEO II.V data.

To estimate the systematic error due to the mass range used in the
maximum likelihood fit ($\pm$40 MeV/c$^2$) a study is performed where
the mass interval is varied and the lifetime recalculated. A variety of
mass intervals are used in the study including narrower intervals (e.g. $\pm$20 MeV/c$^2$),
wider intervals (e.g. $\pm$60 MeV/c$^2$) and asymmetric intervals 
(e.g. $-$60,+40 MeV/c$^2$). From the results of this study a systematic error
of $\pm$10 fs is assigned due to the mass region used in the maximum likelihood fit.  
 
The final measured $\Xi^{+}_{c}$ lifetime value (without systematic
error) is $502.6 \pm 47.3$ fs.  The total systematic uncertainty of
$\pm$18 fs is obtained by adding all the contributions listed in
Table~\ref{tab:mlsystematic} in quadrature.
	
\section{Summary}

A new measurement of the $\Xi^{+}_{c}$ lifetime, 
$\tau_{\Xi^{+}_{c}}=503 \pm 47 ({\rm stat.}) \pm 18 ({\rm syst.})$ fs, 
has been made using the
CLEO II.V detector and 9.0 fb$^{-1}$ of integrated luminosity.
This is the first $\Xi^{+}_{c}$ lifetime measurement from an $e^+e^-$ experiment.
Many of the
contributions to the systematic error in this measurement are different from those of
fixed-target experiments.
This result is somewhat higher
than the current world average, $330^{+60}_{-40}$ fs. We can combine this result
with the recent CLEO II.V measurement~\cite{lbc_cleo} of the $\Lambda^+_c$ lifetime,
$\tau_{\Lambda^{+}_{c}}=179.6 \pm 6.9 ({\rm stat.}) \pm 4.4 ({\rm syst.})$ fs
to obtain $\tau_{\Xi^{+}_{c}}/\tau_{\Lambda^{+}_{c}}=2.8\pm0.3$ (statistical error
only). This measured ratio is higher than the expectation, $\sim1.3$, from the
$1/m_c$ expansion calculation of Ref.~\cite{BBD}.  

\section{Acknowledgments} 
We gratefully acknowledge the effort of the CESR staff in providing us with
excellent luminosity and running conditions.
This work was supported by 
the National Science Foundation,
the U.S. Department of Energy,
the Research Corporation,
the Natural Sciences and Engineering Research Council of Canada
and the Texas Advanced Research Program.

\end{document}